\documentclass[letterpaper,11pt]{article}
\usepackage{hyperref}
\usepackage{amsmath}
\usepackage{amsfonts}
\usepackage{amssymb}
\usepackage{color}
\usepackage{mathrsfs}
\usepackage{latexsym}
\usepackage{physics}
\usepackage{cite}
\usepackage{graphicx}
\usepackage{float}
\usepackage{multirow}
\DeclareMathAlphabet{\mathpzc}{OT1}{pzc}{m}{it}
\usepackage{calligra}
\DeclareMathAlphabet{\mathcalligra}{T1}{calligra}{m}{n}
\title{Trans-Planckian Effects in Warm Inflation}
\author{H. Bouzari Nezhad$^1$, F. Shojai$^{1,2}$,\\$^1$Department of Physics, University of Tehran,\\Tehran, Iran.\\$^2$Foundations of Physics Group, School of Physics,\\Institute for Research in Fundamental Sciences (IPM),\\Tehran, Iran.\\}
\date{}
\begin{document}
\maketitle
\begin{abstract}
We study the effect of a non-trivial vacuum prescription on warm inflation observables, namely the power spectrum of the comoving curvature perturbation. Non-trivial choice of vacuum, can provide information about trans-Planckian physics. Traditionally, the initial condition for inflation is chosen to be the usual Bunch-Davies vacuum. Another more reasonable choice of vacuum is the so-called $\alpha$-vacua. Because the duration of inflation is not infinite, as it is assumed in the Bunch-Davies case, imposing the initial condition at infinite past is not sensible and one must utilize another vacuum prescription. In this paper, working in the slow-roll regime during warm inflation, the initial condition for inflaton fluctuations is imposed at finite past, i.e. the $\alpha$-vacua. We show that this non-trivial vacuum prescription results in oscillatory correction to the comoving curvature power spectrum, which is scale dependent both in amplitude and frequency. Having obtained this scale dependent power spectrum, we consider its late time footprints and compare our results with observational data and other proposed models for the comoving curvature power spectrum.
\end{abstract}
\section{Introduction}
\label{sec:Introduction}
Inflationary scenario \cite{Linde:2005ht,Liddle:2000cg} has been introduced in the 1980s for solving the shortcomings \cite{Guth:1980zm,Linde:1983gd,Starobinsky:1980te,Sato:1980yn} of the standard Big Bang cosmology. In the old version of inflation, which is called the (super-)cold inflation, a scalar field or inflaton is responsible for the very rapid expansion of the Universe. The minimum amount of inflation which will solve the old problems of the standard Big Bang cosmology is about 60 $e$-folds. The period of inflationary phase should not last forever because under this circumstance the Universe becomes extremely dilute which is not compatible with observations. Therefore, there must be a mechanism that terminates the inflationary period. Furthermore, since after the end of inflation the hot Big Bang explosion must happen, this mechanism must fill the Universe with relativistic particles and thus should account for the \textit{reheating} of the Universe.\\
In the traditional cold inflation scenario, it is assumed that the couplings between the inflaton and other fields is relevant only during the reheating period. In the other words, it is supposed that, the produced particles due to these couplings quickly dilute because of the rapid expansion of the Universe and therefore have no effect on the dynamics or consequences of inflation. But, if inflation happens at the Grand Unification scale, i.e. $10^{15}-10^{16}$ GeV, and we suppose that only one part in $10^{20}$ of the inflation's potential energy density is changed into radiation, the resulting temperature of the Universe will be on the order of $10^{11}$ GeV. The triumph of the Big Bang nucleosynthesis is based on the fact that the reheating temperature after inflation should be below $10^9$ GeV. Therefore, it is obvious that the effect of this radiation bath at such high temperature can not be ignored \cite{Albrecht:1982mp,Dolgov:1982th,Abbott:1982hn,Bassett:2005xm}.\\
In order to avoid this mysterious reheating period one can distinguish the model of warm inflation, in which, during the period of inflation the Universe is filled by a self-interacting radiation field and inflaton. In this interesting scenario, the expansion of the Universe is driven by inflaton, while simultaneously, the decay of inflaton into radiation produces the thermal bath \cite{Berera:1995ie,Berera:1996fm,Berera:1996nv,Taylor:2000ze,Hall:2003zp}. One important difference between warm inflation and the usual cold inflation is the primordial fluctuations that are necessary for large scale structures (LSSs) formation. These initial fluctuations for cold inflation are due to quantum effects, while in warm inflation, the thermal fluctuations have an essential role for the formation of LSSs \cite{Moss:1985wn,Berera:1995wh,Berera:1999ws,Herrera:2017qux,Motaharfar:2017dxh}.\\
Inflationary paradigm not only solve the conceptual problems of the standard Big Bang cosmology, but also provides the explanation for the origin of LSSs of the Universe \cite{Mukhanov:1981xt}. In the other words, the seeds of galaxies and cluster of galaxies that we observe today is the fluctuations of inflaton field that are generated during inflation. This mechanism is based on the fact that the fluctuation modes that are well inside the Hubble radius at early times, become larger than the Hubble radius at the end of inflation. Eventually, these fluctuations leave imprints on the cosmic microwave background (CMB) radiation which can be observed today.\\
However, if inflation lasts only slightly more than 70 $e$-foldings (in models which inflation starts near to the Grand Unification scale), then the wavelengths of the fluctuations that are currently inside the Hubble radius, were smaller than the Planck scale at the onset of inflation \cite{Brandenberger:2012aj}. Since the physics on scales smaller than the Planck scale is yet unknown, to understand the origin and evolution of the fluctuation modes at very early times, some new physics is required. This problem is called the trans-Planckian problem of inflationary cosmology.\\
There are several approaches to attacking the trans-Planckian problem of inflationary cosmology; see \cite{Brandenberger:2000wr,Martin:2000xs,Niemeyer:2000eh,Danielsson:2002kx,Kaloper:2002uj,Burgess:2002ub,Burgess:2003zw} for original related works. In \cite{Danielsson:2002kx}, which is the subject of the present work, the trans-Planckian problem has been discussed from the viewpoint of the choice of vacuum. Traditionally, the initial condition for inflaton fluctuations is chosen to be the so-called Bunch-Davies vacuum \cite{Bunch:1978yq}. In the Bunch-Davies vacuum prescription, it is implicitly assumed that the infinite past is accessible and at this moment the spacetime resembles Minkowskian. The critical idea behind the approach that has been introduced in \cite{Danielsson:2002kx} is that since the duration of inflation is finite, therefore, imposing the initial condition at infinite past may lead to inaccurate results. Furthermore, one can not follow a given fluctuation mode to infinitely small scales, because the current known physics is restricted to the scales larger than the Planck scale.\\
In \cite{Danielsson:2002kx}, the inaccessibility of the Minkowskian vacuum at infinite past is justified by introducing a momentum cutoff $\Lambda$, and it is assumed that the evolution of the fluctuation modes begin when the physical momentum $k$ related to a given fluctuation mode satisfy $k=a(t_i)\Lambda$, where $a(t_i)$ is the scale factor evaluated at the initial time $t_i$, when the initial condition is imposed. In this work we use this interesting vacuum choice, which is known as the $\alpha$-vacua, in warm inflation. In the slow-roll regime, which we will focus on it in this paper, we define $z_i\equiv-k\tau_i=\Lambda/H(t_i)$, where $\tau_i$ is the initial conformal time and $H(t_i)$ is the Hubble parameter at the initial time $t_i$.\\
The remainder of this paper is organized as follows: In section \ref{sec:Warm Inflation Scenario}, the basics and definitions of warm inflation will be introduced. Section \ref{sec:Perturbation Theory} is devoted to the subject of perturbation theory in the framework of warm inflation. After calculating the inflaton and metric fluctuations, these quantities will be used to obtain the comoving curvature perturbation. The initial condition for inflaton fluctuations will be imposed at finite past, which is the definition of the $\alpha$-vacua. The power spectrum of the comoving curvature perturbation will be considered and we show that there is a significant and observable difference in using the Bunch-Davies vacuum and the $\alpha$-vacua. In section \ref{sec:Planck Likelihood Analysis and Comparison with Observational Data} we perform the Planck Likelihood Analysis and compare our result with observational data. We end up with conclusion in section \ref{sec:Conclusion}.\\
Throughout this paper we use $M^2_{Pl}=(8\pi G)^{-1}$ and mostly positive metric signature, $(-,+,+,+)$. Furthermore, the superscripts $BD$ and $\alpha$ refers to the Bunch-Davies vacuum and the $\alpha$-vacua, respectively.
\section{Warm Inflation Scenario}
\label{sec:Warm Inflation Scenario}
Before proceeding further, let us review the main aspects and definitions that we will encounter in warm inflation scenario.
\subsection{The Dynamics of Warm Inflation}
In warm inflation, in order to avoid the reheating period, the inflatons must decay into a significant amount of particles during the inflationary period, and thus we will have a smooth transition from the inflationary phase into the radiation dominated stage. Usually, it is assumed that these produced particles create a thermal gas of radiation with temperature $T$. The production of the radiation can be considered by introducing a damping term in the equation of motion for inflaton, which will be
\begin{eqnarray}\label{EofM}
\ddot\phi+(3H+\Upsilon)\dot\phi+\partial_\phi V=0
\end{eqnarray}
where $\phi(t)$ is the inflaton field, $H(t)$ is the Hubble parameter, $\Upsilon(\phi,T)$ is the friction coefficient, and $V(\phi,T)\equiv V_0(\phi)+V_R(\phi,T)$ is the thermodynamic potential. Here $V_0$ is the potential energy density of inflatons which also appears in the usual cold inflation scenario and $V_R$ is the potential energy density of free radiation field plus a thermal correction due to the interaction of inflaton with radiation. Therefore in the Friedmann equations, we must include the total energy density and the pressure \cite{Hall:2003zp}
\begin{eqnarray}\label{totalenergydensityandpressure}
\begin{split}
&\rho=\rho_\phi+\rho_R=\frac{1}{2}\dot\phi^2+(V_0+V_R)+sT\\&
p=p_\phi+p_R=\frac{1}{2}\dot\phi^2-(V_0+V_R)
\end{split}
\end{eqnarray}
in which $s$ and $T$ are the entropy density and the temperature of the radiation, respectively\footnote{In classical thermodynamics \cite{Pathria:2011}, the Helmholtz free energy $A$ is defined by $A=E-TS$, where $E$, $T$, and $S$ are the energy, temperature, and entropy of the system, respectively. These quantities together with pressure $p$, and the volume $L^3$ of the system are related through the Euler's relation, i.e. $E=TS-pL^3$. Comparison of these two expressions gives $A=-pL^3$. This relation simply states that the pressure of the system is the minus of the Helmholtz free energy density. Differentiating $A=E-TS$ leads to
\[
S=-\left(\frac{\partial A}{\partial T}\right)_{E,S}
\]
}. Subtracting the two equations in (\ref{totalenergydensityandpressure}) and using the known facts that $\rho_\phi-p_\phi=2V_0$ and \footnote{Defining the entropy and energy densities by $s=S/L^3$ and $\rho=E/L^3$, respectively, we obtain
\[
E-TS=A=-pL^3\Rightarrow sT=\rho+p
\]} $\rho_R+p_R=sT$, we conclude that $V_R=-p_R$. As mentioned before, the pressure of radiation is the minus of the Helmholtz free energy density $A/L^3$, and therefore $V_R=A/L^3$. The calculation of the Helmholtz free energy of a radiation field has been done in the framework of finite temperature quantum field theory and the result for $V_R(\phi,T)$ is \cite{Linde:1978px,Bellac:2011kqa}
\begin{eqnarray}\label{thermodynamicpotential}
V_R(\phi,T)=-\frac{1}{90}\pi^2g_\star T^4+\frac{1}{2}(\delta m_T)^2\phi^2+...
\end{eqnarray}
where $g_\star(T)$ is the effective degrees of freedom of radiation field and $\delta m_T(\phi,T)$ represents thermal corrections. To obtain a relation for entropy density, it is necessary to notice that in warm inflation, the decay of inflaton field is responsible for the production of entropy. In general relativity, the conservation of energy-momentum tensor reads
\begin{eqnarray}\label{energy-momentmconservationI}
\dot\rho+3H(\rho+p)=0
\end{eqnarray}
Substituting from Eq. (\ref{totalenergydensityandpressure}) yields
\begin{eqnarray}\label{energy-momentmconservationII}
\dot\phi\ddot\phi+\dot\phi\partial_\phi(V_0+V_R)+\dot T\partial_T(V_0+V_R)+\dot sT+s\dot T+3H(\dot\phi^2+Ts)=0
\end{eqnarray}
Using the equation of motion for inflaton in (\ref{EofM}), $\partial_TV_0=0$, and $\partial_TV_R=\partial_T(A/L^3)=-s$ in Eq. (\ref{energy-momentmconservationII}), we arrive at
\begin{eqnarray}\label{energy-momentmconservationIII}
\dot sT+3HsT=\Upsilon\dot\phi^2
\end{eqnarray}
Eq. (\ref{energy-momentmconservationIII}) simply states that the entropy increases due to the increase of the dissipation rate but decreases as the Universe expands. For highly relativistic particles we have $p_R=(1/3)\rho_R$ which together with Eq. (\ref{totalenergydensityandpressure}) and the explanations after it, gives
\begin{eqnarray}\label{forradiationI}
\rho_R=\frac{1}{30}\pi^2g_\star T^4,~~V_R=-\frac{1}{3}\rho_R,~~sT=\frac{4}{3}\rho_R,~~\dot s=\frac{1}{T}\dot\rho_R
\end{eqnarray}
where the thermal correction term in Eq. (\ref{thermodynamicpotential}) has been ignored. The first equation in (\ref{forradiationI}) is nothing but the usual Stefan-Boltzmann law. Because by this approximation, i.e. $p_R=(1/3)\rho_R$, we are treating the inflaton field and the radiation as uncorrelated, this result is not far from what was expected \cite{Taylor:2000ze}. Taking into account the relations in Eq. (\ref{forradiationI}), the differential equation for the entropy density, Eq. (\ref{energy-momentmconservationIII}), can be transformed into the following differential equation for the radiation energy density
\begin{eqnarray}\label{forradiationII}
\dot\rho_R+4H\rho_R=\Upsilon\dot\phi^2
\end{eqnarray}
This differential equation for the radiation energy density is more intuitive than that one for the entropy energy density, i.e. Eq. (\ref{energy-momentmconservationIII}). Eq. (\ref{forradiationII}) indicates that the radiation energy density increases due to the dissipation rate $\Upsilon$ and decreases because of the Hubble expansion of the Universe. The factor of four in front of $H$ in comparison with three in Eq. (\ref{energy-momentmconservationIII}) states that the radiation energy density receives an extra decrease which is due to the redshift of the radiation \cite{Berera:2006xq}. Collecting our results, we obtain the following equations that completely determine the dynamics of warm inflation
\begin{eqnarray}\label{warminflationequations}
\begin{split}
&H^2=\frac{1}{3M_{Pl}^2}(\frac{1}{2}\dot\phi^2+V_0+\rho_R)\\&
\ddot\phi+3H(1+r)\dot\phi+\partial_\phi V_0=0\\&
\dot\rho_R+4H\rho_R-\Upsilon\dot\phi^2=0
\end{split}
\end{eqnarray}
where parameter $r=\Upsilon/3H$ is the ratio of the thermal damping to the expansion damping and for warm inflation $r>1$. For the given functions $V_0(\phi)$ and $\Upsilon(\phi,T)$, the above equations completely determine the dynamics of warm inflation, that is the time evolution of the scale factor $a(t)$, the inflaton field $\phi(t)$, and the radiation energy density $\rho_R(t)$. With $\rho_R(t)$ in hand, one can straightforwardly calculate $s(t)$ and $T(t)$ by using relations in Eq. (\ref{forradiationI}). However this procedure is complicated and usually the slow-roll approximation is used.
\subsection{The Slow-Roll Parameters}
One can simplify the warm inflation equations in (\ref{warminflationequations}) if the following conditions are met
\begin{eqnarray}\label{approximation}
\frac{\dot\phi^2/2+\rho_R}{V_0}\ll1,~~\frac{\ddot\phi}{3Hr\dot\phi}\ll1,~~\frac{\dot\rho_R}{4H\rho_R}\ll1
\end{eqnarray}
where we have also restricted ourselves to high dissipation regime, i.e. $r\gg1$. By these conditions, we can neglect terms of the highest order in time derivatives from Eq. (\ref{warminflationequations}) and the warm inflation equations become
\begin{eqnarray}\label{srwarminflationequations}
\begin{split}
&H^2-\frac{1}{3M_{Pl}^2}V_0\approx0\\&
3Hr\dot\phi+\partial_\phi V_0\approx0\\&
4H\rho_R-\Upsilon\dot\phi^2\approx0
\end{split}
\end{eqnarray}
As in the case of the usual cold inflation, the validity of the slow-roll approximation is controlled by the slow-roll parameters. There are four leading order slow-roll parameters in warm inflation
\begin{eqnarray}\label{slow-rollparameters}
\begin{split}
&\epsilon=\frac{M_{Pl}^2}{2}(\frac{\partial_\phi V_0}{V_0})^2,~~\eta=M_{Pl}^2\frac{\partial_\phi\partial_\phi V_0}{V_0},\\&\beta=M_{Pl}^2\frac{\partial_\phi V_0}{V_0}\frac{\partial_\phi\Upsilon}{\Upsilon},~~\delta=\frac{T\partial_T\partial_\phi V}{\partial_\phi V}
\end{split}
\end{eqnarray}
The first two of these, are the usual cold inflation slow-roll parameters whereas the second two are the characteristic of warm inflation. Using the first inequality in Eq. (\ref{approximation}) and the slow-roll conditions in (\ref{srwarminflationequations}), we conclude that 
\begin{eqnarray}\label{fsrcI}
\epsilon\ll2r
\end{eqnarray}
where we have assumed that $r\gg1$. On the other hand, by differentiating the first equation in (\ref{srwarminflationequations}) with respect to the time and using the second equation to eliminate $\dot\phi$, we obtain the following useful relation \footnote{For a nearly exponential inflation, we must have  $|\dot H|/{H^2}<1$ which requires $\epsilon<r$.}
\begin{eqnarray}\label{fsrcII}
-\frac{\dot H}{H^2}=\frac{\epsilon}{r}
\end{eqnarray}
By using Eqs. (\ref{srwarminflationequations}), (\ref{slow-rollparameters}), (\ref{fsrcII}), the second and third inequalities of (\ref{approximation}) yield
\begin{eqnarray}\label{satsrc}
\begin{split}
&\frac{\ddot\phi}{3Hr\dot\phi}=\frac{\beta-\eta}{3r^2}\ll1\Longrightarrow\beta-\eta\ll3r^2\\&
\frac{\dot\rho_R}{4H\rho_R}=\frac{\beta}{4r}-\frac{\eta}{2r}+\frac{\epsilon}{4r}\ll1\Longrightarrow\frac{\beta}{4}-\frac{\eta}{2}+\frac{\epsilon}{4}\ll r
\end{split}
\end{eqnarray}
Now, the slow-roll conditions in Eq. (\ref{approximation}), translate into constraints on the slow-roll parameters $\epsilon$, $\eta$, and $\beta$, i.e. Eqs. (\ref{fsrcI}) and (\ref{satsrc}) \cite{Hall:2003zp}. Using these three equations, one can easily conclude that the slow-roll approximation holds as long as $\epsilon$, $\eta$, and $\beta$ are much smaller than $r$. In the next section, when we consider the perturbation theory in warm inflation, Eqs. (\ref{fsrcI}) and (\ref{satsrc}) will be very beneficial.
\section{Perturbation Theory}
\label{sec:Perturbation Theory}
Perhaps, the most important triumph of inflation is that it provides a convincing explanation for large scale structures of the Universe which we observe today. In the usual cold inflation scenario the origin of these large scale structures is usually considered to be the quantum fluctuations of the inflaton field. Rather than quantum effects, in warm inflation, the origin of these large scale structures is the thermal effects. This means that thermal fluctuations in the radiation, will cause the inflaton to fluctuates. Therefore it is necessary to calculate the inflaton perturbations due to the thermal fluctuations of the radiation. Then we evaluate the effect of inflaton fluctuations on the metric. Finally, the metric fluctuations will be used to calculate the power spectrum of the comoving curvature perturbation.
\subsection{Inflaton Perturbations}
Adding a small perturbation to background inflaton field, i.e. $\phi\rightarrow\phi+\delta\phi$, and using the unperturbed inflaton equation in (\ref{EofM}) we arrive at the following equation for inflaton perturbations
\begin{eqnarray}\label{inflatonequationI}
\delta\ddot\phi+(3H+\Upsilon)\delta\dot\phi-(\frac{\vec\nabla}{a})^2\delta\phi=0
\end{eqnarray}
where we have ignored $(\partial_\phi\partial_\phi V_0)\delta\phi$, because according to Eq. (\ref{slow-rollparameters}) this term is second order in small parameters. Until now, Eq. (\ref{inflatonequationI}) does not include the effect of radiation that produces thermal perturbations. The influence of radiation can be included by adding an extra noise term $\xi(\vec x,t)$ to the right hand side of Eq. (\ref{inflatonequationI}). This procedure is called the Schwinger-Keldysh approach to non-equilibrium field theory which describes the interaction of a scalar field with radiation \cite{Schwinger:1960qe,Keldysh:1964ud}. After adding the extra noise term, the resulting equation which describes the evolution of a scalar field in the presence of radiation is just a Langevin equation \cite{Gleiser:1993ea,Calzetta:1986cq}
\begin{eqnarray}\label{Langevinequation}
\delta\ddot\phi(\vec x,t)+(3H+\Upsilon)\delta\dot\phi(\vec x,t)-(\frac{\vec\nabla}{a})^2\delta\phi(\vec x,t)=\xi(\vec x,t)
\end{eqnarray}
Doing the Fourier transform over $\vec x$ we obtain the following equation for the evolution of perturbation modes $\delta\phi_k(t)$
\begin{eqnarray}\label{LangevinFourierI}
\delta\ddot\phi_k(t)+(3H+\Upsilon)\delta\dot\phi_k(t)+(\frac{k}{a})^2\delta\phi_k(t)=\xi_k(t)
\end{eqnarray}
where $\delta\phi_k(t)$ and $\xi_k(t)$ are the Fourier transforms of $\delta\phi(\vec x,t)$ and $\xi(\vec x,t)$, respectively. Usually, it will be assumed that the noise term has a Gaussian distribution and is white. Under these assumptions, the correlation function of the noise is Markovian \cite{Gleiser:1993ea,Hall:2003zp}
\begin{eqnarray}\label{noisecorrelation}
\langle\xi_k(t)\xi_{-k'}(t')\rangle_\beta=2\Upsilon T(\frac{2\pi}{a})^3\delta^{(3)}(k-k')\delta(t-t')
\end{eqnarray}
where the subscript $\beta$ means that averaging over the noise should be done. The delta functions in Eq. (\ref{noisecorrelation}) guarantee that the different modes that the noise is composed of them, are perfectly well uncorrelated and this is the definition of the white noise. It is straightforward to solve Eq. (\ref{LangevinFourierI}) under the condition (\ref{noisecorrelation}). In the appendix we have shown that after making a change of variable from $t$ to $z(t)=k/\left(a(t)H(t)\right)$, the solution of Eq. (\ref{noisecorrelation}) is
\begin{eqnarray}\label{solutionII}
\delta\phi_k(z)=\frac{\pi}{2H^2}\int_{z}^{z_i}\left[J_\nu(z)Y_\nu(y)-Y_\nu(z)J_\nu(y)\right]\frac{z^{\nu}}{y^{\nu+1}}\xi_k(y)dy
\end{eqnarray}
where $J_\nu(z)$ and $Y_\nu(z)$ are the Bessel and Neumann functions of order $\nu\equiv(3/2)(1+r)$, respectively. As we stated in the appendix, for the slow-roll inflation $z=-k\tau$. Since we impose the initial condition at finite past, i.e. the $\alpha$-vacua, the upper bound of the integral is set to $z_i=-k\tau_i$. For the usual Bunch-Davies vacuum, it goes to $\infty$. The power spectrum for inflaton perturbations is defined via
\begin{eqnarray}\label{powerspectrumI}
\langle\delta\phi_k(z)\delta\phi_{-k'}(z)\rangle_\beta=(\frac{2\pi}{k})^3\mathcal{P}_{\phi\phi}(k)\delta(k-k')
\end{eqnarray}
Substituting from Eq. (\ref{solutionII}) we obtain
\begin{eqnarray}\label{powerspectrumII}
\begin{split}
\langle\delta\phi_k(z)\delta\phi_{-k'}(z)\rangle_\beta&=(\frac{\pi}{2H^2})^2\int_{z}^{z_i}\int_{z}^{z_i}\left[J_\nu(z)Y_\nu(y)-Y_\nu(z)J_\nu(y)\right]\frac{z^{\nu}}{y^{\nu+1}}\times\\&\left[J_\nu(z)Y_\nu(x)-Y_\nu(z)J_\nu(x)\right]\frac{z^{\nu}}{x^{\nu+1}}\langle\xi_k(y)\xi_{-k'}(x)\rangle_\beta dxdy
\end{split}
\end{eqnarray}
The correlation function that appears in the above equation, can be rewritten using Eq. (\ref{noisecorrelation}). To perform the integration, it is necessary to write the integration variables $dx$ and $dy$ in terms of $t$. To do this, we note that actually we have $x(t_x)=k/\left(a(t_x)H(t_x)\right)$ and $x(t_y)=k/\left(a(t_y)H(t_y)\right)$. Using Eq. (\ref{chainruleII}) in the appendix, it can be easily seen that in the slow-roll approximation, $dx=-xHdt_x$ and $dy=-yHdt_y$. Replacing these new integration variables into Eq. (\ref{powerspectrumII}) and then using Eq. (\ref{noisecorrelation}), we obtain
\begin{eqnarray}\label{solutionIII}
\begin{split}
\langle\delta\phi_k(z)\delta\phi_{-k'}(z)\rangle_\beta=&(\frac{\pi}{2H^2})^2\int_{z}^{z_i}\left[J_\nu(z)Y_\nu(y)-Y_\nu(z)J_\nu(y)\right]^2\frac{z^{2\nu}}{y^{{2\nu+2}}}\times\\&(-yH)2\Upsilon T(\frac{2\pi}{a})^3\delta^{(3)}(k-k')dy
\end{split}
\end{eqnarray}
Comparison of Eqs. (\ref{powerspectrumI}) and (\ref{solutionIII}), leads to the following expression for the power spectrum of inflaton fluctuations
\begin{eqnarray}\label{powerspectrumIII}
\mathcal{P}_{\phi\phi}^{\alpha}(k)=-\frac{\pi^2}{2}\Upsilon Tz^{2\nu}\int_{z}^{z_i}\left[J_\nu(z)Y_\nu(y)-Y_\nu(z)J_\nu(y)\right]^2\frac{dy}{y^{2\nu-2}}
\end{eqnarray}
Eq. (\ref{powerspectrumIII}) is the exact expression for the power spectrum of inflaton fluctuations. As usual, we restrict ourselves to the times where a given mode with physical wavelength $k$ has crossed the horizon. This means that we have $k/(aH)\ll1$ and thus the arguments of the Bessel and the Neumann functions, $z$, are some positive number much smaller than unity. For small real arguments, these functions have the following asymptotic forms \cite{Arfken:2012}
\begin{eqnarray}\label{BesselandNeumannAsyptotic}
J_\nu(z)\approx\frac{1}{\nu!}(\frac{z}{2})^\nu,~~~Y_\nu(z)\approx-\frac{\Gamma(\nu)}{\pi}(\frac{z}{2})^{-\nu}
\end{eqnarray}
Substituting these asymptotic expressions into Eq. (\ref{powerspectrumIII}), we conclude that for small values of $z$, the first term in the parentheses of (\ref{powerspectrumIII}) vanishes. Also we can safely set the lower limit of the above integrals to zero. Therefore the power spectrum will be obtained as follows
\begin{eqnarray}\label{powerspectrumV}
\begin{split}
\mathcal{P}_{\phi\phi}^{\alpha}(k)=&-\frac{\pi^2}{8}\Upsilon Tz^{2\nu}Y_\nu^2(z)z_i^3\Gamma(\nu+\frac{1}{2})\times\\&_{2}\tilde F_3(\frac{3}{2},\frac{1}{2}+\nu;\frac{5}{2},1+\nu,1+2\nu;-z_i^2)\\
\end{split}
\end{eqnarray}
where $_{p}\tilde F_q(a_1,...,a_p;b_1,...,b_q;z)$ is the regularized generalized hypergeometric function which is defined as
\begin{eqnarray}\label{}
_{p}\tilde F_q(a_1,...,a_p;b_1,...,b_q;z)=\frac{_{p}F_q(a_1,...,a_p;b_1,...,b_q;z)}{\Gamma(b_1)...\Gamma(b_q)}
\end{eqnarray}
in which $_{p}F_q(a_1,...,a_p;b_1,...,b_q;z)$ is the generalized hypergeometric function \cite{Arfken:2012}. Setting $z_i=\Lambda/H(t_i)$, it follows that if we treat $\Lambda$ as the energy scale in which some new physics emerges, i.e. Planck scale, $\Lambda/H(t_i)\gg 1$ \cite{Danielsson:2002kx}. Therefore we can expand the regularized generalized hypergeometric function in Eq. (\ref{powerspectrumV}) for large $z_i^2$. At the same time, we can approximate the Neumann function by Eq. (\ref{BesselandNeumannAsyptotic}). The resulting expression for the power spectrum is
\begin{eqnarray}\label{powerspectrumVII}
\mathcal{P}_{\phi\phi}^{\alpha}(k)=-\sqrt{\frac{\pi\Upsilon H}{4}}T\left[1-\frac{\sqrt{2}}{\pi}\Gamma(2\nu)\left(\frac{\Lambda}{H(t_i)}\right)^{-2\nu+1}\cos(\frac{2\Lambda}{H(t_i)}-\pi\nu)\right]
\end{eqnarray}
where we have substituted $z=k/(aH)$, $z_i=\Lambda/H(t_i)$, $\nu\approx\Upsilon/(2H)$, and we have only shown the leading order correction. In obtaining the above result we have used the following properties of the Gamma functions \cite{Arfken:2012}
\begin{eqnarray}\label{Gammafp}
2^{2\nu}\Gamma(\nu)=4\sqrt{\pi}\frac{\Gamma(2\nu-1)}{\Gamma(\nu)-\frac{1}{2}},~~~\frac{\Gamma(\nu+a)}{\Gamma(\nu+b)}\approx \nu^{a-b}
\end{eqnarray}
in which the first relation is known as the duplication formula and the second one holds for large values of $\nu$, high dissipation regime. It should be mentioned that if instead of considering the leading term in the expansion of Neumann function, Eq. (\ref{BesselandNeumannAsyptotic}), the next terms are included, we will get additional corrected terms to the power spectral function which are proportional to the powers of $k/a$ and will be ignorable if $(k/a)^2\ll H\Upsilon$. This inequality defines a new physical moment, called the freeze-out moment \cite{Hall:2003zp}, which is the analogous of the horizon crossing moment for the cold inflation, i.e. $k/a=H$, and defined by the following relation\footnote{The freeze-out moment happens sooner than the horizon crossing moment, because earlier times correspond to greater values of $z$.}
\begin{eqnarray}\label{freeze-out}
\frac{k}{a}=\sqrt{H\Upsilon}
\end{eqnarray}
In order to clarify the definition of the freeze-out moment, we should keep in mind that the evolution of the comoving inflaton fluctuation modes during warm inflation are divided into three different regimes which are thermal noise, expansion, and curvature fluctuations. The transmissions between these regimes are called the freeze-out and the horizon crossing moments, respectively. After the freeze-out moment, the gravitational perturbations stop evolving \cite{Berera:2006xq}.\\
It should be remembered that without the square brackets in Eq. (\ref{powerspectrumVII}), the above result is the familiar power spectrum resulted from the Bunch-Davies vacuum, which have been obtained in \cite{Hall:2003zp} and also in the context of the stochastic inflation approach in \cite{Ramos:2013nsa}. In \cite{Matsuda:2009eq}, the evolution of curvature perturbations during warm inflation has been discussed from the viewpoint of the $\delta N$ formalism.
\subsection{Cosmological Perturbations}
Now, we consider the metric fluctuations. First of all, we note that the metric fluctuations appear in the form of the scalar, vector, and tensor fluctuations. The tensor perturbations are believed to be the origin of the gravitational waves and we do not consider them here. As it is shown in \cite{Weinberg:2008zzc}, vector perturbations decay as the inverse squared of the scale factor and have not played a significant role in cosmology. So, only the scalar perturbations are important and we consider them here (for more details see \cite{Weinberg:2008zzc,Mukhanov:2005sc}). The basic idea is that we write the spacetime metric, $g_{\mu\nu}$, as the usual  Friedmann–Lemaître–Robertson–Walker (FLRW) background, $\bar g_{\mu\nu}$, plus a small perturbation $\delta g_{\mu\nu}$. This means that
\begin{eqnarray}\label{perturbedmetric}
g_{\mu\nu}=\bar g_{\mu\nu}+\delta g_{\mu\nu}=\begin{pmatrix}
-1+2\Phi&0\\
0&a^2(1+2\Phi)\delta_{ij}
\end{pmatrix}
\end{eqnarray}
where $\Phi(\vec{x},t)$ is the scalar perturbation. The perturbed metric in Eq. (\ref{perturbedmetric}) has been written in the Newtonian gauge and by the assumption that the anisotropic part of the stress tensor is absent \cite{Weinberg:2008zzc}. The evolution of the scalar perturbation $\Phi(\vec{x},t)$ can be found by perturbing the Einstein field equations. The non-zero components of the perturbed Einstein tensor can be easily calculated \cite{Weinberg:2008zzc}. The energy-momentum tensor has two contributions from the inflaton field and the radiation. One can easily show that in the slow-roll approximation, the dominant contribution is due to the inflaton field and the contribution of the radiation can be ignored. To clear this point, we note that in the slow-roll approximation, the total energy density of the inflaton is dominated by its potential energy, which gives $\rho_\phi\approx V_0$. Then, using Eq. (\ref{srwarminflationequations}) and the definitions of $\epsilon$ and $r$ we conclude that
\begin{eqnarray}\label{proofI}
\rho_R\approx\frac{\epsilon}{2r}\rho_\phi
\end{eqnarray}
which is precisely what we have claimed. The energy-momentum tensor for the inflaton field is $T_{\mu\nu}=(\partial_\mu\phi)(\partial_\nu\phi)+g_{\mu\nu}\mathcal{L}$, where $\mathcal{L}=-(1/2)g^{\mu\nu}(\partial_\mu\phi)(\partial_\nu\phi)$\\$-V_0(\phi)$ is its Lagrangian. Perturbing this energy-momentum tensor, the non-zero components of the Einstein field equations read
\begin{eqnarray}\label{perturbedEinsteinII}
\begin{split}
3H\dot\Phi_k+(\frac{k}{a})^2\Phi_k=&\frac{1}{2M_{Pl}^2}(\dot\phi\delta\phi_k-2\Phi_kV_0+\delta V_0)\\
\dot\Phi_k+H\Phi_k=&-\frac{1}{2M_{Pl}^2}\dot\phi\delta\phi_k\\
\ddot\Phi_k+4H\dot\Phi_k+4\frac{\ddot a}{a}\Phi_k+2H^2\Phi_k=&-\frac{1}{2M_{Pl}^2}(\dot\phi\delta\dot\phi_k-2\Phi_kV_0-\delta V_0+2\Phi_k\dot\phi^2)
\end{split}
\end{eqnarray}
where $\Phi_k$ and $\delta\phi_k$ are the Fourier transforms of $\Phi$ and $\delta\phi$, respectively. In practice, it is very hard to solve the set of differential equations in (\ref{perturbedEinsteinII}). The good news is that from the viewpoint of observational cosmology, one is not interested in the explicit form of the function $\Phi_k(t)$. Comparison of the inflationary models with recent observational data is done by considering the comoving curvature perturbation $\mathcal{R}$, which is a gauge independent measure of the deviation from FLRW metric. In Fourier space, this quantity is $\mathcal{R}_k=\Phi_k+H\delta\phi_k/\dot\phi$. As mentioned before at the end of subsection 3.1, the gravitational perturbation $\Phi_k$ becomes a constant at large scales. Therefore, the scalar perturbation is given by the second equation in (\ref{perturbedEinsteinII}) where the first term in the left hand side is neglected. Thus, the comoving curvature perturbation $\mathcal{R}_k$ would be
\begin{eqnarray}\label{cspI}
\begin{split}
\mathcal{R}_k=\Phi_k+\frac{H\delta\phi_k}{\dot\phi}\approx\left(-\frac{1}{2M_{Pl}^2}\frac{\dot\phi}{H}+\frac{H}{\dot\phi}\right)\delta\phi_k&=\left(-\frac{\epsilon}{r^2}+1\right)\frac{H}{\dot\phi}\delta\phi_k\\&\approx\frac{H}{\dot\phi}\delta\phi_k
\end{split}
\end{eqnarray}
where we have used Eqs. (\ref{srwarminflationequations}) and (\ref{slow-rollparameters}) alongside with this fact that the slow-roll parameter $\epsilon$ is much smaller than $r$. Similar to Eq. (\ref{powerspectrumI}), one can define the power spectrum for comoving curvature perturbation $\mathcal{R}_k$ as follows
\begin{eqnarray}\label{pscspI}
\langle\mathcal{R}_k\mathcal{R}_{-k'}\rangle_\beta=(\frac{2\pi}{k})^3\mathcal{P}_\mathcal{R}(k)\delta(k-k')
\end{eqnarray}
Substituting from Eq. (\ref{powerspectrumVII}), the power spectrum for comoving curvature perturbation would be
\begin{eqnarray}\label{pscspII}
\begin{split}
&\mathcal{P}_\mathcal{R}^{\alpha}(k)=\frac{H^2}{\dot\phi^2}\mathcal{P}_{\phi\phi}^{\alpha}(k)=\\
&-\sqrt{\frac{\pi\Upsilon}{4}}\frac{H^{5/2}T}{\dot\phi^2}\left[1-\frac{\sqrt{2}}{\pi}\Gamma(2\nu)\left(\frac{\Lambda}{H(t_i)}\right)^{-2\nu+1}\cos(\frac{2\Lambda}{H(t_i)}-\pi\nu)\right]
\end{split}
\end{eqnarray}
Without the square brackets, the above result is the standard expression for the power spectrum of the comoving curvature perturbation when the Bunch-Davies vacuum was used \cite{Hall:2003zp}
\begin{eqnarray}\label{B-DI}
\mathcal{P}_\mathcal{R}^{BD}(k)=-\sqrt{\frac{\pi\Upsilon}{4}}\frac{H^{5/2}T}{\dot\phi^2}
\end{eqnarray}
In order to show that the correction term in Eq. (\ref{pscspII}) is scale dependent, we differentiate the initial condition $k=a(t_i)\Lambda$ with respect to $k$
\begin{eqnarray}\label{icI}
\frac{dt_i}{dk}=\frac{1}{\Lambda a(t_i)H(t_i)}
\end{eqnarray}
Using this and Eq. (\ref{fsrcII}) evaluated at the initial time $t_i$, we obtain
\begin{eqnarray}\label{icII}
\frac{dH(t_i)}{H(t_i)}=-\frac{\epsilon(t_i)}{r(t_i)}\frac{dk}{k}
\end{eqnarray}
Integrating the above equation and ignoring the $k$ dependence of $\epsilon(t_i)/r(t_i)$, we get
\begin{eqnarray}\label{icIII}
H(t_i)=H_{f}(\frac{k}{k_{f}})^{-\epsilon/r}
\end{eqnarray}
where $H_f$ is the value of Hubble parameter at the time when the first scale, $k_f$, satisfies the initial condition \cite{Nezhad:2018qqj}. Now let us to evaluate the power spectrum of the comoving curvature perturbation at the horizon crossing moment, $t_k$.
For this purpose, we must evaluate all quantities outside the square brackets in (\ref{pscspII}) at the horizon crossing moment, $t_k$. Differentiating the horizon crossing condition, $k/a(t_k)=H(t_k)$, with respect to $k$ gives
\begin{eqnarray}\label{hcI}
\frac{dt_k}{dk}=\frac{1}{a(t_k)[H^2(t_k)+\dot H(t_k)]}
\end{eqnarray}
Using this and Eq. (\ref{fsrcII}) at the horizon crossing moment, we obtain
\begin{eqnarray}\label{hcII}
\frac{k}{H(t_k)}\frac{dH(t_k)}{dk}=-\frac{\epsilon(t_k)/r(t_k)}{1-\epsilon(t_k)/r(t_k)}
\end{eqnarray}
Integrating the above equation assuming that $\epsilon(t_k)/r(t_k)$ is small and the $k$ dependence of it is ignorable, gives
\begin{eqnarray}\label{hcIII}
H(t_k)=H_{l}(\frac{k}{k_{l}})^{-\epsilon/r}
\end{eqnarray}
where $H_l$ is the value of Hubble parameter at the time when the last scale $k_l$ has left the horizon. To obtain the $k$ dependence of $\Upsilon(t_k)$, we use Eqs. (\ref{srwarminflationequations}), (\ref{slow-rollparameters}), (\ref{hcI}), and $r=\Upsilon/3H$. Thus
\begin{eqnarray}\label{hcIV}
\frac{k}{\Upsilon(t_k)}\frac{d\Upsilon(t_k)}{dk}=-\frac{\beta(t_k)}{r(t_k)}\frac{1}{1-\epsilon(t_k)/r(t_k)}
\end{eqnarray}
which leads to
\begin{eqnarray}\label{hcV}
\Upsilon(t_k)=\Upsilon_l(\frac{k}{k_l})^{-\beta/r}
\end{eqnarray}
Using Eqs. (\ref{satsrc}) and (\ref{hcI}), we obtain the following differential equation for $\dot\phi(t_k)$
\begin{eqnarray}\label{hcVI}
\frac{k}{\dot\phi(t_k)}\frac{d\dot\phi(t_k)}{dk}=\frac{\beta(t_k)-\eta(t_k)}{r(t_k)}\frac{1}{1-\epsilon(t_k)/r(t_k)}
\end{eqnarray}
which after integration gives
\begin{eqnarray}\label{hcVII}
\dot\phi(t_k)=\dot\phi_l(\frac{k}{k_l})^{(\beta-\eta)/r}
\end{eqnarray}
Finally, we obtain the $k$ dependence of $T(t_k)$. Note that using the third and fourth relations in (\ref{forradiationI}) and Eq. (\ref{satsrc}), we obtain two useful relations
\begin{eqnarray}\label{hcVIII}
\frac{\dot\rho_R}{4H\rho_R}=\frac{\dot s}{3Hs}=\frac{\beta}{4r}-\frac{\eta}{2r}+\frac{\epsilon}{4r},~~~~~\frac{1}{3}\frac{\dot s}{s}=\frac{\dot T}{T}
\end{eqnarray}
Using the above equations and Eq. (\ref{hcI}) we get
\begin{eqnarray}\label{hcIX}
\frac{k}{T(t_k)}\frac{dT(t_k)}{dk}=\left[\frac{\beta(t_k)}{4r(t_k)}-\frac{\eta(t_k)}{2r(t_k)}+\frac{\epsilon(t_k)}{4r(t_k)}\right]\frac{1}{1-\epsilon(t_k)/r(t_k)}
\end{eqnarray}
Integrating the above equation, we find
\begin{eqnarray}\label{hcX}
T(t_k)=T_l(\frac{k}{k_l})^{\beta/4r-\eta/2r+\epsilon/4r}
\end{eqnarray}
In Eqs. (\ref{hcV}), (\ref{hcVII}), and (\ref{hcX}), $\Upsilon_l$, $\dot\phi_l$, and $T_l$ are the value of the corresponding quantities at the time when the last scale $k_l$ has left the horizon. It should be remembered that in obtaining these results, we have assumed that $\epsilon(t_k)/r(t_k)$ is small and the $k$ dependence of it is ignorable. Putting all these results into Eq. (\ref{pscspII}), the power spectrum of the comoving curvature perturbation would be
\begin{eqnarray}\label{pscspIII}
\begin{split}
&\mathcal{P}_\mathcal{R}^{\alpha}\left(k=a(t_k)H(t_k)\right)=\mathcal{A}_{\alpha}(\frac{k}{k_l})^{-\frac{1}{r}\left[\frac{9}{4}(\epsilon+\beta)-\frac{3}{2}\eta\right]}\times\\&\left[1-\frac{\sqrt{2}}{\pi}\Gamma(2\nu)(\frac{\Lambda}{H_f})^{-2\nu+1}(\frac{k}{k_f})^{(-2\nu+1)\epsilon/r}\cos(\frac{2\Lambda}{H_f}(\frac{k}{k_f})^{\epsilon/r}-\pi\nu)\right]
\end{split}
\end{eqnarray}
where we have defined $\mathcal{A}_{\alpha}\equiv-\sqrt{\pi}\Upsilon_l^{1/2}H_l^{5/2}T_l\dot\phi_l^{-2}/2$. The scalar spectral index $n_s$ is defined through the relation
\begin{eqnarray}\label{ssiI}
n_s-1=\frac{d\ln\mathcal{P}_\mathcal{R}(k)}{d\ln k}
\end{eqnarray}
Substituting from Eq. (\ref{pscspIII}) we get
\begin{eqnarray}\label{ssiII}
\begin{split}
n_s^{\alpha}-1=&-\frac{1}{r}\left[\frac{9}{4}(\epsilon+\beta)-\frac{3}{2}\eta\right]\\&+\frac{2\sqrt{2}}{\pi}\Gamma(2\nu)\frac{\epsilon}{r}(\frac{\Lambda}{H_f})^{-2\nu+2}(\frac{k}{k_f})^{(-2\nu+2)\epsilon/r}\sin(\frac{2\Lambda}{H_f}(\frac{k}{k_f})^{\epsilon/r}-\pi\nu)
\end{split}
\end{eqnarray}
where we have only shown the leading order correction. The first term in the above equation is the standard scalar spectral index for warm inflation when the initial condition is chosen to be the Bunch-Davies vacuum \cite{Hall:2003zp}
\begin{eqnarray}\label{BDssi}
n_s^{BD}-1=-\frac{1}{r}\left[\frac{9}{4}(\epsilon+\beta)-\frac{3}{2}\eta\right]
\end{eqnarray}
The second term in Eq. (\ref{ssiII}) shows that the leading correction due to the $\alpha$-vacua has an oscillatory behavior with scale dependent amplitude and frequency.
\section{Planck Likelihood Analysis and Comparison with Observational Data}
\label{sec:Planck Likelihood Analysis and Comparison with Observational Data}
\begin{figure}[h!]
\begin{center}
\includegraphics[scale=0.7]{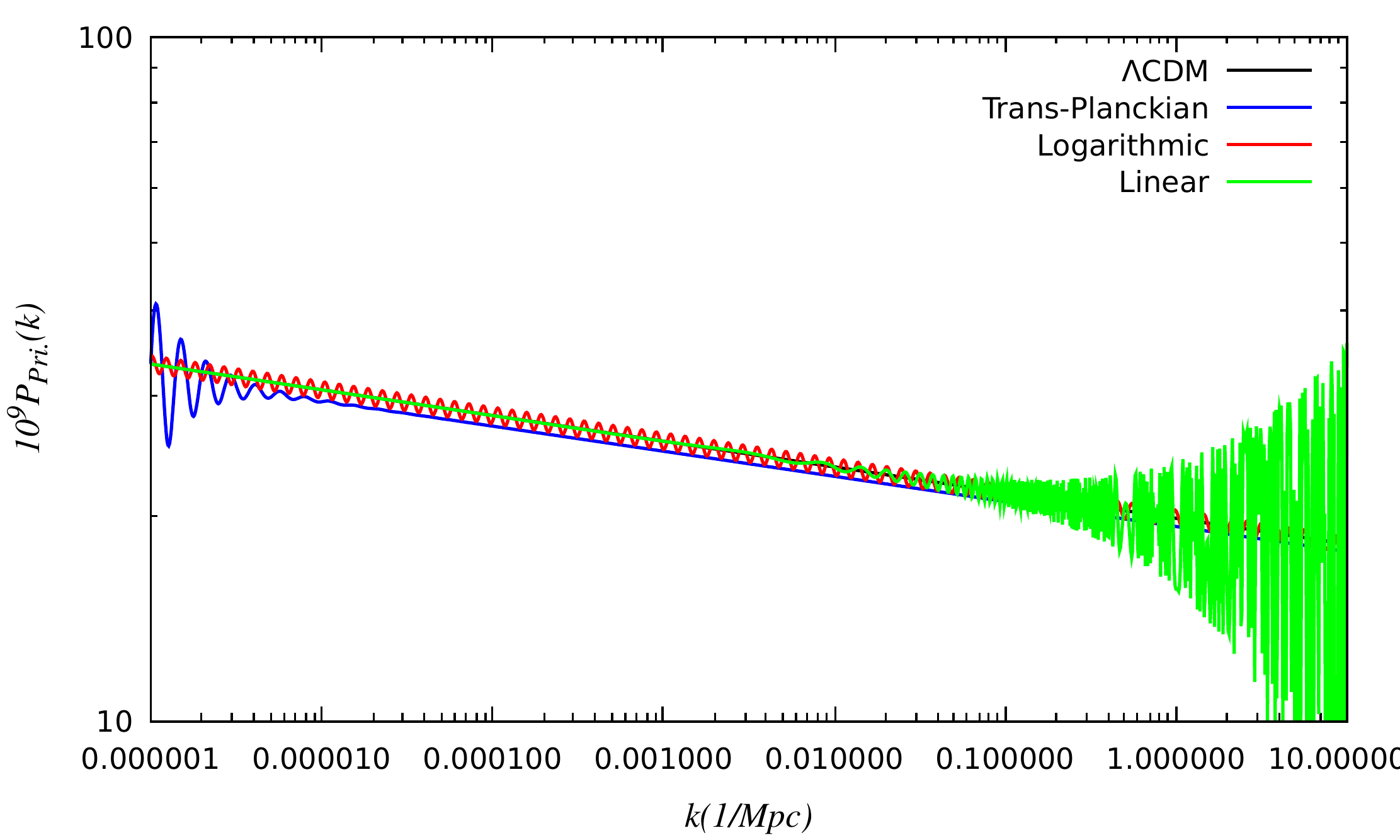}
\caption{Dimensionless primordial power spectra for different models. Black, blue, red, and green lines are for the $\Lambda$CDM Eq. (\ref{LCDM}), trans-Planckian Eq. (\ref{pscspIV}), logarithmic Eq. (\ref{logarithmic}), and linear Eq. (\ref{linear}), respectively. These power spectra have been plotted by using the best-fit values for their free parameters in Table (\ref{tab:table1}) and Table (\ref{tab:table2}). The $\Lambda$CDM power spectrum coincides with the linear power spectrum when the oscillatory behavior of this model not yet started.}
\label{fig:01}
\end{center}
\end{figure}

\begin{figure}[h!]
\begin{center}
\includegraphics[scale=0.7]{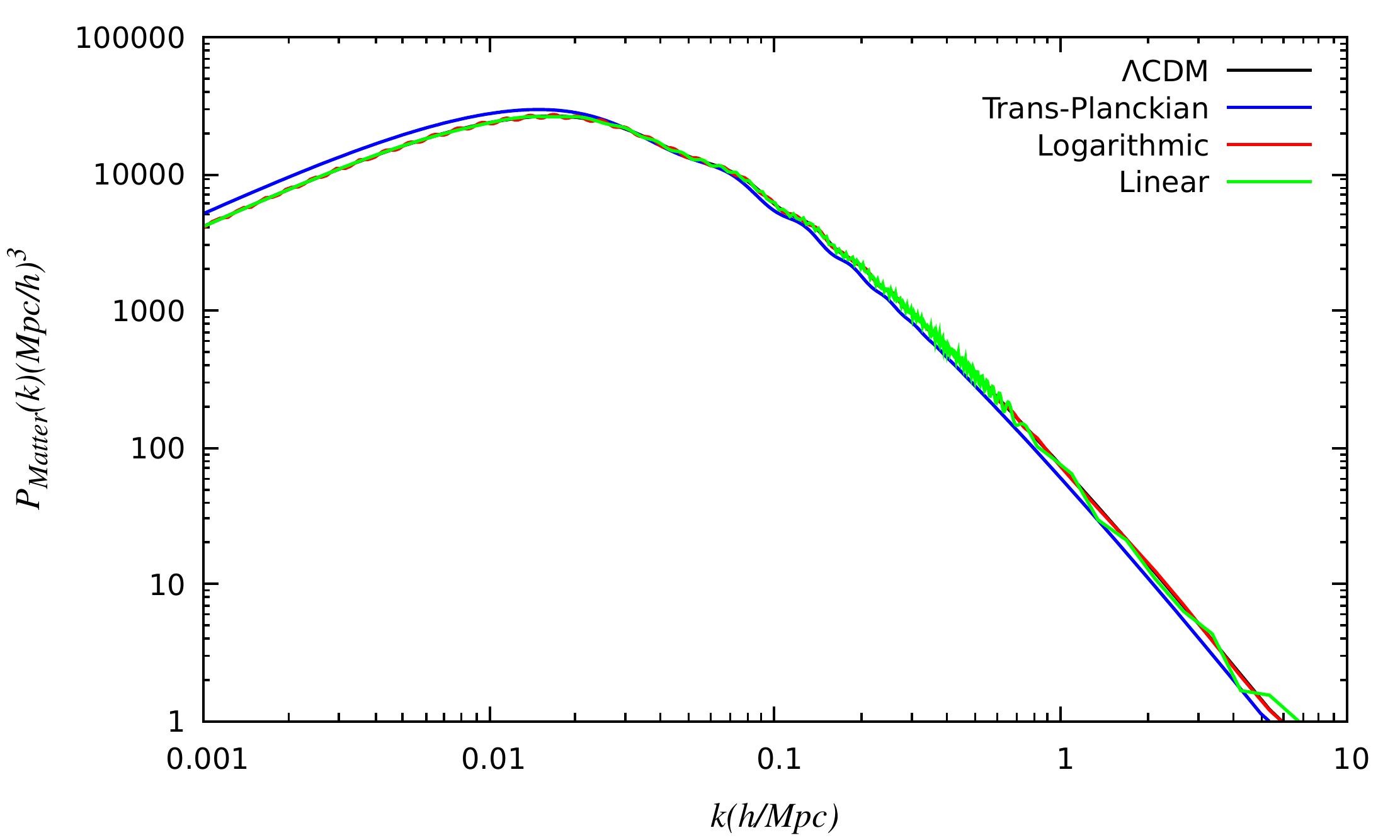}
\caption{Total matter power spectrum at redshift ${\mathcal Z}=0$ for different models using the best-fit values in Table (\ref{tab:table1}) and Table (\ref{tab:table2}). The black, blue, red, and green lines are for the $\Lambda$CDM Eq. (\ref{LCDM}), trans-Planckian Eq. (\ref{pscspIV}), logarithmic Eq. (\ref{logarithmic}), and linear Eq. (\ref{linear}), respectively.}
\label{fig:02}
\end{center}
\end{figure}

\begin{figure}[h!]
\begin{center}
\includegraphics[scale=0.7]{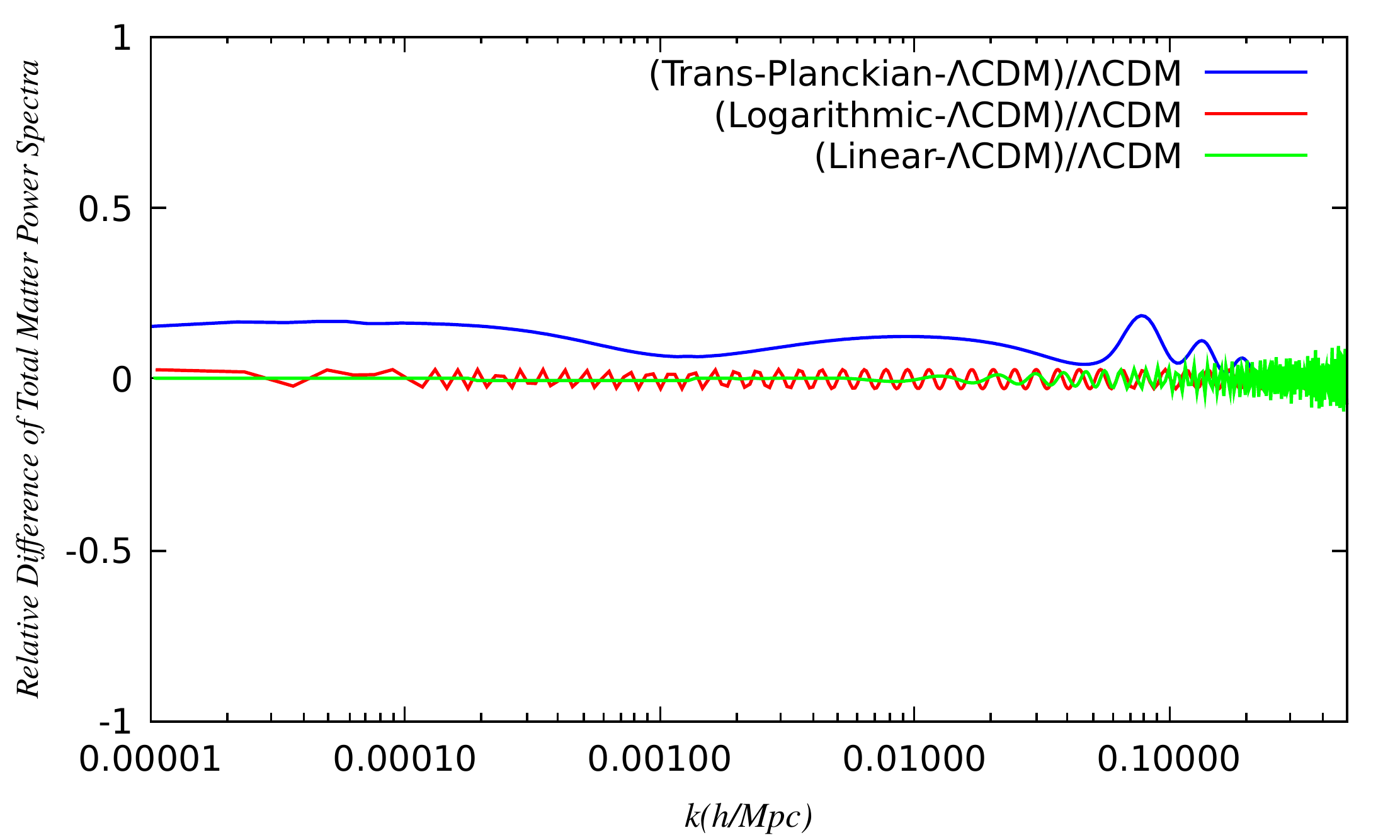}
\caption{Relative difference of the total matter power spectra with $\Lambda$CDM model at ${\mathcal Z}=0$ using the best-fit values in Table (\ref{tab:table1}) and Table (\ref{tab:table2}).}
\label{fig:03}
\end{center}
\end{figure}

\begin{figure}[h!]
\begin{center}
\includegraphics[scale=0.7]{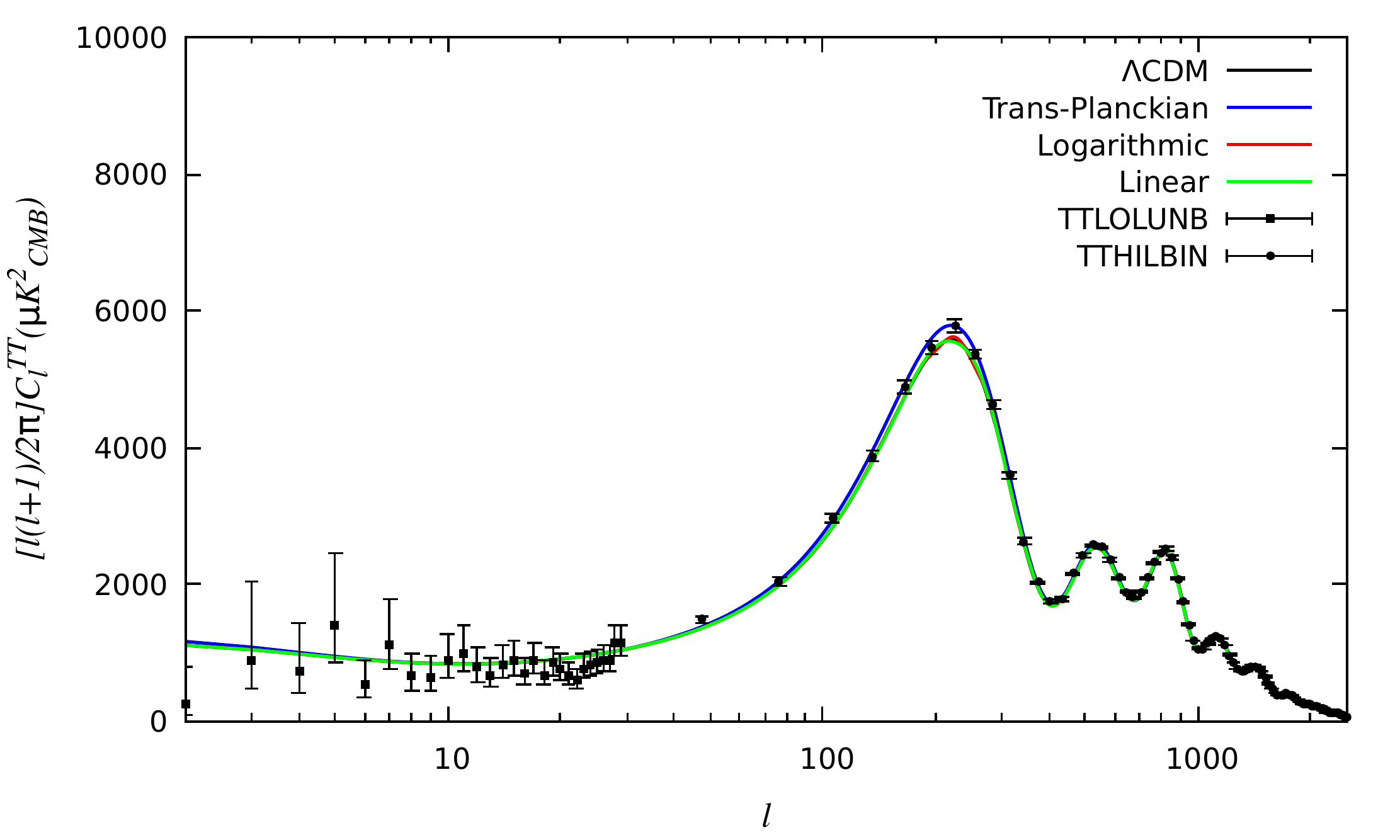}
\caption{Lensed CMB temperature anisotropy. Black, blue, red, and green lines are for the $\Lambda$CDM Eq. (\ref{LCDM}), trans-Planckian Eq. (\ref{pscspIV}), logarithmic Eq. (\ref{logarithmic}), and linear Eq. (\ref{linear}), respectively. Black squares are the Planck low $l$ unbinned (TTLOLUNB) results. Black circles are the Planck high $l$ binned (TTHILBIN) results. The Planck data have been plotted at 1$\sigma$ confidence level. The interesting point in these plots is the better matching of our trans-Planckian model with the Planck data.}
\label{fig:04}
\end{center}
\end{figure}

\begin{figure}[h!]
\begin{center}
\includegraphics[scale=0.2]{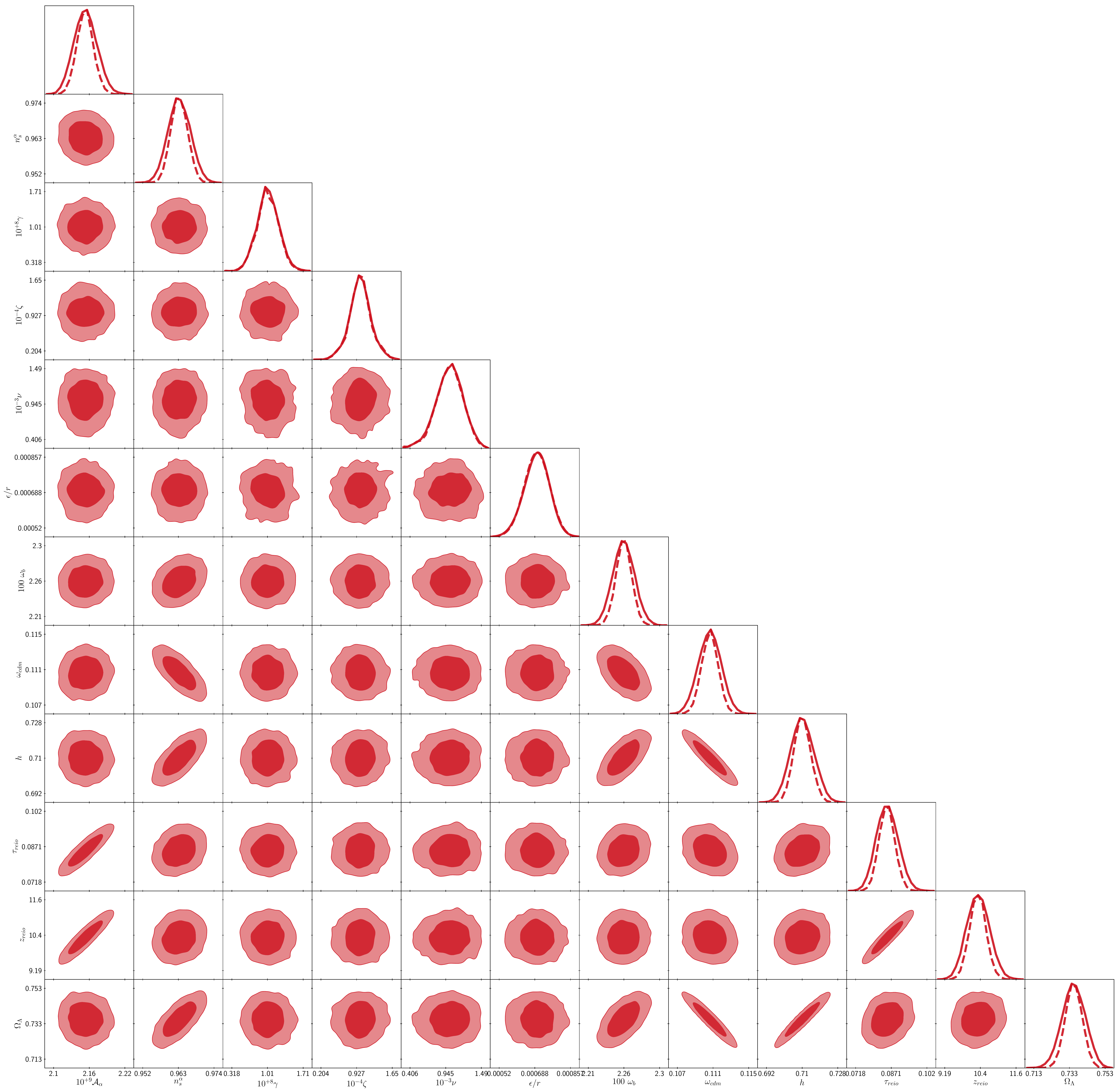}
\caption{Triangle plot for our model parameters. Off diagonal plots represent 1$\sigma$ (dark contours) and 2$\sigma$ (bright contours) confidence limits for the variation of two sets of model parameters. Diagonal plots are the marginalized posterior probability distributions (solid lines) and mean likelihood (dashed lines).}
\label{fig:05}
\end{center}
\end{figure}

In this section we perform the Planck Likelihood Analysis on the trans-Planckian power spectrum in Eq. (\ref{pscspIII}), and compare the late time observables, such as the total matter power spectrum and the CMB temperature anisotropies, with the Planck data. To do this job we use the CLASS (Cosmic Linear Anisotropy Solving System) code \cite{Blas:2011rf} alongside with the Monte Python \cite{Audren:2012wb}. Also our sampling method is the Nested Sampling through MultiNest \cite{Feroz:2008xx}. We begin by Eq. (\ref{pscspIII}) and express this power spectrum in the following form
\begin{eqnarray}\label{pscspIV}
\begin{split}
\mathcal{P}_\mathcal{R}^{\alpha}\left(k=a(t_k)H(t_k)\right)=&\mathcal{A}_{\alpha}(\frac{k}{k_l})^{n_s^{\alpha}-1}\times\\&\left[1-\gamma(\frac{k}{k_f})^{(-2\nu+1)\epsilon/r}\cos(2\zeta(\frac{k}{k_f})^{\epsilon/r}-\pi\nu)\right]
\end{split}
\end{eqnarray}
where we have defined
\begin{eqnarray}\label{comparisonI}
-\frac{1}{r}\left[\frac{9}{4}(\epsilon+\beta)-\frac{3}{2}\eta\right]\equiv n_s^{\alpha}-1,~~\frac{\sqrt{2}}{\pi}\Gamma(2\nu)(\frac{\Lambda}{H_f})^{-2\nu+1}\equiv\gamma,~~\frac{\Lambda}{H_f}\equiv\zeta
\end{eqnarray}
In order to do this comparison more clearly, we also compare our result with two other power spectra which has deviations from the standard power spectrum in Eq. (\ref{B-DI}). Similar to our work, non-Bunch-Davies initial conditions \cite{Martin:2000xs,Danielsson:2002kx,Bozza:2003pr} or the axion monodromy model \cite{Silverstein:2008sg,Kobayashi:2012kc,McAllister:2008hb} lead to the logarithmic modification of the comoving curvature power spectrum
\begin{eqnarray}\label{logarithmic}
\mathcal{P}_{\mathcal{R}}^{log}(k)=\mathcal{P}_{\mathcal{R}}^{0}(k)\left\{1+\mathcal{A}_{log}\cos\left[\omega_{log}\ln(\frac{k}{k_{*}})+\phi_{log}\right]\right\}
\end{eqnarray}
where the best-fit values for the free parameters of this power spectrum are given in Table (\ref{tab:table1}). Consideration of the boundary effective field theories \cite{Jackson:2013vka,Meerburg:2013dla} results in the linear oscillation of the power spectrum of comoving curvature perturbation
\begin{eqnarray}\label{linear}
\mathcal{P}_{\mathcal{R}}^{lin}(k)=\mathcal{P}_{\mathcal{R}}^{0}(k)\left[1+\mathcal{A}_{lin}(\frac{k}{k_{*}})^{n_{lin}}\cos\left(\omega_{lin}\frac{k}{k_{*}}+\phi_{lin}\right)\right]
\end{eqnarray}
in which the best-fit values of the free parameters are as in Table (\ref{tab:table1}). For completeness sake, we write Eq. (\ref{B-DI}) as
\begin{eqnarray}\label{LCDM}
\mathcal{P}_\mathcal{R}^{BD}(k)\equiv\mathcal{A}_s(k/k_*)^{n_s^{BD}-1}
\end{eqnarray}
where the best-fit values of its parameters are given in Table (\ref{tab:table1}). In what follows we refer to this power spectrum as the $\Lambda$CDM model. Also, it should be mentioned that during our comparison we set $k_l=0.05Mpc^{-1}$ and $k_f=0.01Mpc^{-1}$.
\begin{table}[h!]
\begin{center}
\caption{The best-fit values of the free parameters of the $\Lambda$CDM Eq. (\ref{LCDM}), logarithmic Eq. (\ref{logarithmic}), and linear Eq. (\ref{linear}) primordial power spectra \cite{Ade:2015lrj}.}
\label{tab:table1}
\begin{tabular}{lcc}
\noalign{\hrule\vskip 3pt}
Model & Parameter & Best-fit \\
\noalign{\vskip 3pt\hrule\vskip 3pt}
\multirow{2}{*}{$\Lambda$CDM} & $\mathcal{A}_s$ & $2.215\times10^{-9}$ \\
& $n_s^{B-D}$ & $0.9624$ \\
\noalign{\vskip 3pt\hrule\vskip 3pt}
\multirow{3}{*}{Logarithmic} & $\mathcal{A}_{log}$ & $0.0278$ \\
& $\log_{10}\omega_{log}$ & $1.51$ \\
& $\phi_{log}/2\pi$ & $0.634$ \\
\noalign{\vskip 3pt\hrule\vskip 3pt}
\multirow{4}{*}{Linear} & $\mathcal{A}_{lin}$ & $0.0292$ \\
& $n_{lin}$ & $0.662$ \\
& $\log_{10}\omega_{lin}$ & $1.73$ \\
& $\phi_{lin}/2\pi$ & $0.554$ \\
\noalign{\hrule\vskip 3pt}
\end{tabular}
\end{center}
\end{table}
The best-fit values of our trans-Planckian power spectrum in Eq. (\ref{pscspIV}) and the other cosmological parameters are given in Table (\ref{tab:table2}). In this table, $\omega_b$, $\omega_{cdm}$, and $\Omega_{\Lambda}$ are the baryon, cold dark matter, and dark energy contributions, respectively, $h\equiv10^{-2}H_0$ is the present value of the Hubble parameter in units of $kms^{-1}Mpc^{-1}$, $\tau_{reio}$ and $z_{reio}$ are the conformal time and the redshift that the reionization occurs, respectively.
\begin{table}[h!]
\begin{center}
\caption{Best-fit values for the trans-Planckian power spectrum in Eq. (\ref{pscspIV}).}
\label{tab:table2}
\begin{tabular}{lcccc}
\noalign{\hrule\vskip 3pt}
Parameter & best-fit & mean$\pm\sigma$ & 95\% lower & 95\% upper \\
\noalign{\vskip 3pt\hrule\vskip 3pt}
$10^{+9}\mathcal{A}_{\alpha}$ & $2.149$ & $2.158_{-0.021}^{+0.019}$ & $2.12$ & $2.196$ \\
\noalign{\vskip 3pt\hrule\vskip 3pt}
$n_s^{B-D}$ & $0.9633$ & $0.9635_{-0.0035}^{+0.0035}$ & $0.9567$ & $0.9705$ \\
\noalign{\vskip 3pt\hrule\vskip 3pt}
$10^{+8}\gamma$ & $1.105$ & $1.023_{-0.21}^{+0.22}$ & $0.5856$ & $1.455$ \\
\noalign{\vskip 3pt\hrule\vskip 3pt}
$10^{-4}\zeta$ & $1.119$ & $1.001_{-0.21}^{+0.21}$ & $0.5286$ & $1.477$ \\
\noalign{\vskip 3pt\hrule\vskip 3pt}
$10^{-3}\nu$ & $1.104$ & $1.002_{-0.19}^{+0.22}$ & $0.5906$ & $1.418$ \\
\noalign{\vskip 3pt\hrule\vskip 3pt}
$\epsilon/r$ & $0.0008451$ & $0.0006968_{-5.1e-05}^{+5.7e-05}$ & $0.0005696$ & $0.0008052$ \\
\noalign{\vskip 3pt\hrule\vskip 3pt}
$100\omega_b$ & $2.253$ & $2.255_{-0.014}^{+0.014}$ & $2.228$ & $2.283$ \\
\noalign{\vskip 3pt\hrule\vskip 3pt}
$\omega_{cdm}$ & $0.1107$ & $0.1108_{-0.0012}^{+0.0012}$ & $0.1084$ & $0.1131$ \\
\noalign{\vskip 3pt\hrule\vskip 3pt}
$h$ & $0.7102$ & $0.7103_{-0.006}^{+0.0057}$ & $0.6989$ & $0.7218$ \\
\noalign{\vskip 3pt\hrule\vskip 3pt}
$\tau_{reio}$ & $0.08354$ & $0.08544_{-0.005}^{+0.0044}$ & $0.0764$ & $0.09469$ \\
\noalign{\vskip 3pt\hrule\vskip 3pt}
$z_{reio}$ & $10.18$ & $10.33_{-0.4}^{+0.37}$ & $9.577$ & $11.1$ \\
\noalign{\vskip 3pt\hrule\vskip 3pt}
$\Omega_{\Lambda}$ & $0.7358$ & $0.7355_{-0.0064}^{+0.0067}$ & $0.7229$ & $0.7484$ \\
\noalign{\hrule\vskip 3pt}
\end{tabular}
$-\ln{\cal L}_\mathrm{min} =0.297269$, minimum $\chi^2=0.5945$ \\
\end{center}
\end{table}
In Fig. (\ref{fig:01}) the power spectrum of comoving curvature perturbation has been plotted for different models. The oscillatory behavior, with scale dependent amplitude and frequency, of our trans-Planckian power spectrum in Eq. (\ref{pscspIV}) for small values of $k$ is obvious.
\\
The difference between the local density and the mean density as a function of scale is called the matter power spectrum and this quantity plays an important role in the consideration of the structure formation in the Universe. This quantity can be expressed as \cite{Dodelson:2003ft}
\begin{eqnarray}\label{matterpowerspectrum}
P_{Matter}(q,{\mathcal Z})=\frac{4}{25}\frac{q^4T^2(q)D^2({\mathcal Z})}{H_0^4\Omega^2_{Matter}}\mathcal{P}_{\mathcal{R}}(q)
\end{eqnarray}
where ${\mathcal Z}$ is the redshift, $T(q)$ is the transfer function, $D({\mathcal Z})$ is the growth function, $H_0$ is the present value of the Hubble parameter, $\Omega_{Matter}$ is the matter density parameter, and $\mathcal{P}_{\mathcal{R}}(q)$ is the primordial power spectrum. In Fig. (\ref{fig:02}) the total matter power spectra for different cases have been shown. For better comparison, the relative difference of these total matter power spectra with the $\Lambda$CDM model have been plotted in Fig. (\ref{fig:03}).
The footprints of primordial power spectrum can be seen in the anisotropies of the CMB. In Fig. (\ref{fig:04}) the lensed CMB temperature anisotropy for different primordial power spectra have been plotted. The preference of our trans-Planckian primordial power spectrum in Eq. (\ref{pscspIV}) over the logarithmic, linear, and $\Lambda$CDM models is that for significant ranges of $l$, our model has better matching with the Planck data. In addition, we have plotted 1D and 2D posterior contours in Fig. (\ref{fig:05})
\section{Conclusion}
\label{sec:Conclusion}
As we mentioned before, since the known physics is restricted to scales which are larger than the Planck scale, one can not follow a given fluctuation mode to scales which are smaller than the Planck scale. Therefore, the usual standard Bunch-Davies vacuum is not very trustworthy and one should resort to another vacuum prescription. In this paper, unlike the Bunch-Davies vacuum, the initial condition for inflaton fluctuations is imposed at finite past, which this kind of vacuum is called the $\alpha$-vacua. In \cite{Danielsson:2002kx,Nezhad:2018qqj}, the effect of the $\alpha$-vacua on the inflationary observables for cold inflation is discussed. In \cite{Danielsson:2002kx}, it has been shown that this effect on the scalar power spectrum is of the order $\Lambda/H$ for de Sitter inflation, while in \cite{Nezhad:2018qqj}, we have shown that if one works in the slow-roll regime from the beginning, this effect is still of the order $\Lambda/H_f$ times a sinousoidal term which indicates the leading order corrections in slow-roll parameters; see Eq. (88) of \cite{Nezhad:2018qqj}. Here, $\Lambda$ is some energy scale in which the evolution of a fluctuation mode begins once $k=a(t_i)\Lambda$, or equivalently $\Lambda$ is an energy scale in which some new physics emerges. Also, $H_f$ is the value of Hubble parameter at the time when the first scale $k_f$ satisfies the initial condition, Eq. (\ref{icIII}). Anyway, as we have shown in \cite{Nezhad:2018qqj}, this non-trivial choice of vacuum leaves observational footprints on the CMB.\\
In the present work, we have considered the effect of this non-trivial vacuum prescription, $\alpha$-vacua, on warm inflationary model. Warm inflation is an alternative for cold inflation with no reheating period. Our final result for the modified power spectrum of the comoving curvature perturbation, i.e. Eq. (\ref{pscspIV}), shows that the correction term in this case is proportional to $(\Lambda/H_f)^{-2\nu+1}$. Usually, one is interested in warm inflationary scenario for high dissipation coefficient, and thus $\nu\gg1$. Also, it is assumed that $\Lambda/H_f\sim10^4$ \cite{Danielsson:2002kx}. Therefore our correction term is small, and due to the factor $(k/k_f)^{(-2\nu+1)\epsilon/r}$, the correction term will be significant only for small values of $k$, where the oscillatory scale dependent amplitude and frequency will be obvious; see Fig. (\ref{fig:01}). The scalar spectral index of warm inflation for the $\alpha$-vacua has been calculated, Eq. (\ref{ssiII}), and we have seen that the correction term is proportional to $(\epsilon/r)(\Lambda/H_f)^{-2\nu+2}$.\\
\renewcommand{\theequation}{A\arabic{equation}}
\setcounter{equation}{0}
\section*{APPENDIX}
In this appendix, we obtain the solution of Eq. (\ref{LangevinFourierI}) with the condition (\ref{noisecorrelation}). We define the new variable $z(t)\equiv k/\left(a(t)H(t)\right)$ and write Eq. (\ref{LangevinFourierI}) in terms of this new variable. Using the chain rule, for any function $f$ we have
\begin{eqnarray}\label{chainruleII}
\begin{split}
&\frac{df}{dt}=-zH\frac{df}{dz}\\&
\frac{d^2f}{dt^2}=z^2H^2\frac{d^2f}{dz^2}+zH^2\frac{df}{dz}
\end{split}
\end{eqnarray}
where we have used Eq. (\ref{fsrcII}) and in the slow-roll approximation we have ignored those terms which are first and higher order in $\epsilon/r$. Thanks to these equations, the differential equation (\ref{LangevinFourierI}) can be written in terms of $z$ as
\begin{eqnarray}\label{LangevinFourierII}
z^2H^2\frac{d^2}{dz^2}\delta\phi_k+zH^2\left[1-(3H+\Upsilon)\frac{1}{H}\right]\frac{d}{dz}\delta\phi_k+z^2H^2\delta\phi_k=\xi_k
\end{eqnarray}
To recast the above differential equation to a more familiar form, we do a transformation as $\delta\phi_k\rightarrow z^\nu\tilde{\delta\phi_k}$, where $\nu\equiv(3/2)(1+r)$. Under this transformation, the differential equation (\ref{LangevinFourierII}) becomes
\begin{eqnarray}\label{Bessel}
z^2\frac{d^2}{dz^2}\tilde{\delta\phi_k}+z\frac{d}{dz}\tilde{\delta\phi_k}+(z^2-\nu^2)\tilde{\delta\phi_k}=\frac{\xi_k}{z^\nu H^2}
\end{eqnarray}
which is the inhomogeneous Bessel equation of order $\nu$. At this point it is more appropriate to write Eq. (\ref{fsrcII}) in terms of the new variable $\mathcal{H}\equiv aH$ and the conformal time $\tau$ which is defined through $d\tau=dt/a$. Therefore Eq. (\ref{fsrcII}) would be
\begin{eqnarray}\label{ctI}
\frac{\epsilon}{r}=1-\frac{\mathcal{H}'}{\mathcal{H}^2}
\end{eqnarray}
which prime means differentiation with respect to the conformal time $\tau$. Integrating the above equation and choosing suitable integration constants, give
\begin{eqnarray}\label{ctII}
\mathcal{H}=-\frac{1}{(1-\epsilon/r)\tau}
\end{eqnarray}
which leads to $\mathcal{H}\approx-1/\tau$ in the slow-roll approximation and thus $z=-k\tau$. The range of conformal time $\tau$ is $-\infty<\tau<0$ and thus we have $0<z<\infty$. This is true only in the case of the usual Bunch-Davies vacuum, where the duration of inflation is infinite and the Minkowskian past is accessible. But, if we demand that the duration of inflation is finite, which is called the $\alpha$-vacua, instead of imposing the initial condition at infinite past, i.e. $\tau\rightarrow-\infty$, we must impose the initial condition at some initial time, say $\tau_i$. Therefore, our problem reduces to solving the differential equation (\ref{LangevinFourierII}) subject to the initial conditions $\tilde{\delta\phi_k}(z_i)=d\tilde\delta\phi_k(z_i)/dz=0$. Using the Green's function method \cite{Arfken:2012}, the solution of Eq. (\ref{LangevinFourierII}) can be expressed as
\begin{eqnarray}\label{Green'sfunctionI}
\tilde{\delta\phi_k}(z)=\int_{0}^{\tau_i}G(z,y)\frac{\xi_k(y)}{y^\nu H^2}dy
\end{eqnarray}
where $G(z,y)$ is the Green's function for this problem. First we note that the homogeneous equation has two linearly independent solutions $\tilde{\delta\phi_k}^{(1)}(z)=J_\nu(z)$ and $\tilde{\delta\phi_k}^{(2)}(z)=Y_\nu(z)$, where $J_\nu(z)$ and $Y_\nu(z)$ are the Bessel function of the first kind and second kind (or the Neumann function), respectively. In the region $z<y$, there are no boundary conditions and we are free to write $G(z,y)=C_1(t)J_\nu(z)+C_2(t)Y_\nu(z)$. On the other hand, in the region $z>y$, the Green's function is $G(z,y)=D_1(t)J_\nu(z)+D_2(t)Y_\nu(z)$. Therefore, we have
\begin{eqnarray}\label{Green'sfunctionII}
G(z,y)=\left\{
\begin{array}{ll}
C_1(y)J_\nu(z)+C_2(y)Y_\nu(z),~~~0<z<y\\\\
D_1(y)J_\nu(z)+D_2(y)Y_\nu(z),~~~y<z<z_i
\end{array}
\right.
\end{eqnarray}
Using the following relations for the derivatives of Bessel functions \cite{Arfken:2012}
\begin{eqnarray}\label{Besselderivatives}
\begin{split}
&\frac{d}{dz}J_\nu(z)=\frac{1}{2}\left[J_{\nu-1}(z)-J_{\nu+1}(z)\right]\\&
\frac{d}{dz}Y_\nu(z)=\frac{1}{2}\left[Y_{\nu-1}(z)-Y_{\nu+1}(z)\right]
\end{split}
\end{eqnarray}
we impose the boundary conditions at $z=z_i$. Doing this, we conclude that $D_1(t)=D_2(t)=0$ and thus $G(z,y)=0$ for $y<z<z_i$. Now for the region $0<z<y$, we impose the continuity of the Green's function and the discontinuity of its first derivative, i.e.
\begin{eqnarray}\label{continuityanddiscontinuityI}
\begin{split}
&G(y_-,y)=G(y_+,y)\\&
\frac{\partial G}{\partial z}(y_+,y)-\frac{\partial G}{\partial z}(y_-,y)=\frac{1}{p(y)}
\end{split}
\end{eqnarray}
where $p(y)=y^2$. The result of applying the above conditions are
\begin{eqnarray}\label{}\label{continuityanddiscontinuityII}
C_1(y)=\frac{\pi}{2y}Y_\nu(y),~~~C_2(y)=-\frac{\pi}{2y}J_\nu(y)
\end{eqnarray}
In obtaining these results, the following relation between the Bessel and Neumann functions has been used \cite{Arfken:2012}
\begin{eqnarray}\label{Bessel&Neumann}
Y_\nu(z)\frac{d}{dz}J_\nu(z)-J_\nu(z)\frac{d}{dz}Y_\nu(z)=\frac{2}{z\pi}
\end{eqnarray}
Therefore, the Green's function for this boundary value problem is
\begin{eqnarray}\label{Green'sfunctionIII}
G(z,y)=\frac{\pi}{2y}\left[J_\nu(z)Y_\nu(y)-Y_\nu(z)J_\nu(y)\right]\theta(y-z)
\end{eqnarray}
where $\theta(y-z)$ is the Heaviside step function. Finally, the solution of Eq. (\ref{LangevinFourierI}) would be
\begin{eqnarray}\label{solutionI}
\tilde{\delta\phi_k}(z)=z^{-\nu}\delta\phi_k(z)=\frac{\pi}{2H^2}\int_{z}^{z_i}\left[J_\nu(z)Y_\nu(y)-Y_\nu(z)J_\nu(y)\right]\frac{\xi_k(y)}{y^{\nu+1}}dy
\end{eqnarray}
\section*{Acknowledgments}
F. Shojai is grateful to the University of Tehran for supporting this work under a grant provided by the university research council.

\end{document}